\documentclass{article}

\usepackage{eus}

\usepackage[boxed]{algorithm}
\usepackage{algorithmic}

\usepackage{graphicx} %  \DeclareGraphicsExtensions{.pdf}

\title{Blind separation of noisy Gaussian stationary sources.\\
  Application to cosmic microwave background imaging.}

\date{}

\author{
  Jean-Fran\c cois Cardoso,  CNRS / ENST - TSI, Paris France
  \\
  Hichem Snoussi, L2S - Supelec, Gif-sur-Yvette, France
  \\
  Jacques Delabrouille, Guillaume Patanchon
  PCC - Coll\`ege de France, Paris, France
}

\def\nbd{{m}}
\def\nbc{{n}}

\def\dom{\mathcal{D}}
\def\diag{\mathrm{diag}}
\def\trace{\mathrm{tr}}
\def\inv{^{-1}}
\def\adj{^\dagger }
\def\inv{^{-1}}

\begin{document}

\maketitle

\begin{abstract}
  
  We present a new source separation method which maximizes the
  likelihood of a model of \emph{noisy} mixtures of stationary,
  possibly Gaussian, independent components.  The method has been
  devised to address an astronomical imaging problem.  It works in the
  spectral domain where, thanks to two simple approximations, the
  likelihood assumes a simple form which is easy to handle (low
  dimensional sufficient statistics) and to maximize (via the EM
  algorithm).

\end{abstract}

\section{SOURCE SEPARATION for ASTRONOMY}

\subsection{Astronomical components}\label{eq:astro}

Source separation consists in recovering components from a set of
observed mixtures.
Component separation is a topic of major interest to the Planck space 
mission, to be launched in 2007 by ESA to map the cosmic microwave 
background (CMB).  The blackbody temperature of this radiation as a 
function of direction on the sky will be measured in $m=10$ different 
frequency channels, corresponding to wavelengths ranging from 
$\lambda=350\,$microns to $\lambda=1\,$cm.  In each channel, the 
temperature map will show not only the CMB contribution but also 
contributions from other sources called \emph{foregrounds}, among 
which Galactic dust emission, emission from very remote (and hence 
quasi point-like) galaxy clusters, and several others.
It is expected that (after some heavy pre-processing), the map built
from the $i$-th channel can be accurately modeled as $y_i(\vec
r)=\sum_{j=1}^{n}a_{ij} s_j(\vec r) + n_i(\vec r)$ where $s_j(\vec r)$
is the spatial pattern for the $j$-th component and $n_i(\vec r)$ is
an additive detector noise.
In other words, cosmologists expect to observe a noisy instantaneous 
(\emph{i.e.} non convolutive) mixture of essentially independent 
components (independence being the consequence of the physically 
distinct origins of the various components).
Even though recovering as cleanly as possible the CMB component is the
primary goal of the mission, astrophysicists are also interested in
the \emph{other} components, in particular for collecting data
regarding the morphology and physical properties of Galactic
foregrounds (dust\ldots) and the distribution of galaxy clusters.

This paper deals with \emph{blind} component separation.  Blindness
means recovering the components without knowing in advance the
coefficients of the mixture.  In practice, this may be achieved by
resorting to the strong but often plausible assumption of mutual
statistical independence between the components.
The motivation for a blind approach is obvious: even though some
coefficients of the mixture may be known in advance with good accuracy
(in particular those related to the CMB), some other components are
less well known or predictable.  It is thus very tempting to run blind
algorithms which do not require \emph{a priori} information about the
mixture coefficients.

\subsection{Blind separation methods}\label{sec:bsm}

Several attempts at blind component separation for CMB imaging have
already been reported.  The first proposal, due to Baccigalupi
\textit{et al.} use a non Gaussian noise-free i.i.d. model for the
components\cite{2000MNRAS.318..769B}, hence following the `standard'
path to source separation.  One problem with this approach is that the
most important component, namely the CMB itself, seems to closely
follow a Gaussian distribution.  It is well known that, in i.i.d.
models, it is possible to accommodate at most one Gaussian component.
It does not seem to be a good idea, however, to use a non Gaussian
model when the main component itself has a Gaussian distribution.

Another reason why the i.i.d. modeling (which is implicit in
`standard' ICA) probably is not appropriate to our application: most
of the components are very much dominated by the low-frequency part of
their spectra.  Thus sample averages taken through the data set tend
not to converge very quickly to their expected values.  This may
explain why Fourier methods, presented below, seem to perform better.

Thus, rather than exploiting the non Gaussianity of (all but one of)
the components, one may think of exploiting their spectral diversity.
A very sensible approach is proposed by Pham: using the Whittle
approximation of the likelihood, he shows that blind separation can be
achieved by jointly diagonalizing spectral covariance matrices
computed over several frequency bands~\cite{pham:eusipco2000}.  This
conclusion however is reached only in the case of noise-free models.
Therefore, it is not appropriate for CMB imaging where a very
significant amount of noise is expected.

In this paper, we follow Pham's approach but we take additive noise
into account, leading to a likelihood function which is no longer a
joint diagonality criterion, thus requiring some new algorithmics.  We
present below the form taken by the EM algorithm when applied to a set
of spectral covariance matrices.  This approach leads to an efficient
algorithm, much faster than the algorithms obtained via the EM
technique in the case of non Gaussian i.i.d. modeling as
in~\cite{EMMG:icassp97} or~\cite{snoussi:MaxEnt2001}.

\subsection{A stationary Gaussian model}

Our method is obtained by starting from a stationary Gaussian model.
For ease of exposition, we assume that the observations are $m$ times
series rather than $m$ images (extension to images is
straightforward).  The $m\times 1$-dimensional observed process
$y(t)=[y_1(t); \ldots; y_m(t)]$ is modeled as
\begin{equation}
  \label{eq:model}
  y(t)=As(t) + n(t)
\end{equation}
where $A$ is an $m\times n$ matrix with linearly independent columns.
The $n$-dimensional source process $s(t)$ (the components) and the
$m$-dimensional noise process $n(t)$ are modeled as real valued,
mutually independent and stationary with spectra $S_s(\nu )$ and
$S_n(\nu )$ respectively.  The spectrum of the observed process then
is
\begin{equation}
  \label{eq:specmody}
  S_y(\nu ) 
  =
  A S_s(\nu ) A\adj + S_n(\nu ) 
  .
\end{equation}
The $\dagger$ superscript denotes transconjugation even though
transposition would be enough almost everywhere in this paper (our
method is easily adapted to deal with complex signals/mixtures).
The assumption of independence between components implies that
$S_s(\nu )$ is a diagonal matrix:
\begin{displaymath}
  [ S_s(\nu ) ]_{ij} = \delta_{ij} P_i(\nu ) 
\end{displaymath}
where $P_i(\nu )$ is the power spectrum of the $i$th source at
frequency $\nu $.
For simplicity, we also assume that the observation noise is
uncorrelated both in time and across sensors:
\begin{equation}
  \label{eq:whitenoise}
  [S_n(\nu )]_{ij} = \delta_{ij} \sigma_i^2
\end{equation}
meaning that the noise spectral density $\sigma_i^2$ on the $i$th
detector does not depend on frequency $\nu $.
In summary the probability distribution of the process is fully
specified by $m\times n$ mixture coefficients, $m$ noise levels and
$n$ power spectra.

\section{THE OBJECTIVE FUNCTION}\label{sec:blind-separ-meth}

Our method boils down to adjusting smoothed versions of the spectral
covariance matrices~(\ref{eq:specmody}) to their empirical estimates.
The estimated parameters are those which give the best match, as
measured by an objective function.  This objective function is
introduced in this section.  In the following section, we show how it
stems from the maximum likelihood principle.

\subsection{Spectral averaging.}

A key feature of our method is that it uses low dimensional statistics
obtained as averages over \emph{spectral domains} in Fourier space.
These Fourier domains simply are frequency bands in the 1D case or are
two-dimensional domains of the Fourier plane when the method is
applied to images.

Consider a partition of the frequency interval $(-\frac 1 2, \frac 1
2)$ into $Q$ domains (bands): $(-\frac 1 2, \frac 1 2) = \cup_{q=1}^Q
\dom_q$ which are required to be symmetric: $f\in\dom_q\Rightarrow
-f\in\dom_q$.
For any function $f(\nu )$ of frequency, denote $\langle f
\rangle_q$ its average over the $q$-th spectral domain when sampled at
multiples of $1/T$:
\begin{equation}
  \label{eq:defspecav}
  \langle f \rangle_q 
  = 
  \frac 1 {w_q}  \sum_{\frac p T \in \dom_q} f \left(\frac p T\right)
  ,\ \ \ \ \ \ \ q=1,\ldots, Q
\end{equation}
where $w_q$ is the number of points in domain $\dom_q$.

\subsection{Spectral statistics}\label{sec:spectral-statistics}

Denoting $Y(\nu )$ the discrete-time Fourier transform of $T$ samples:
\begin{equation}
  \label{eq:dft}
  Y(\nu ) 
  =
  \frac 1 {\sqrt T}  \sum_{t=0}^{T-1} y(t) \exp(-2i\pi \nu  t)
  ,
\end{equation}
the \emph{periodogram} is $\hat S_y(\nu ) = Y(\nu ) Y(\nu )\adj$
and its averaged version is
\begin{equation}
  \label{eq:defsamper}
  \langle \hat  S_y \rangle_q
  = 
  \langle Y(\nu ) Y(\nu )\adj \rangle_q
  .
\end{equation}
Note that $Y(-\nu )=Y(\nu )^*$ for real data so that $ \langle
\hat S_y \rangle_q$ actually is a real valued matrix if $\dom_q$ is a
symmetric domain.

This sample spectral covariance matrix will be our estimate for the
corresponding averaged quantity
\begin{equation}
  \label{eq:modbydom}
    \langle S_y \rangle_q 
    =
    A \langle S_s \rangle_q A\adj
    +
    \langle S_n \rangle_q
\end{equation}
where the equality results from averaging model~(\ref{eq:specmody}).

A key point is that the structure of the model is not affected by
spectral averaging since $\langle S_s \rangle_q $ remains a diagonal
matrix after averaging:
\begin{displaymath}
  \langle S_s \rangle_q 
  =
  \mathrm{diag}\left[  
      \langle P_1 \rangle_q 
      , \ldots,
      \langle P_n \rangle_q 
  \right]
\end{displaymath}
and, of course, $\langle S_n \rangle_q = S_n= \diag(\sigma_1^2,\ldots,
\sigma_m^2) $ still is a constant diagonal matrix.

\subsection{Blind identification via spectral matching}\label{sec:spect-match}

Our proposal for blind identification simply is to match the sample
spectral covariance matrices $\langle \hat S_y \rangle_q$, which
depend on the data, to their theoretical values $\langle S_y
\rangle_q$, which depend on the unknown parameters.
There are $\nbd\times \nbc + Q\times \nbc + \nbd$ of these parameters,
\footnote{ There are actually $\nbc$ redundant parameters since a
  scale factor can be exchanged between each column of $A$ and the
  corresponding power spectra.  } 
collectively referred to as $\theta$:
\begin{equation}\label{eq:deftheta}
  \theta = 
  \left\{
  [A_{ij}]_{i=1,j=1}^{i=\nbd,j=\nbc}  ; \ \
  [\langle P_j \rangle_q ]_{j=1,q=1}^{j=\nbc,q=Q}  ; \ \
  [\sigma_i^2]_{i=1}^{i=\nbd}
  \right\}
  .
\end{equation}
The mismatch between the sample statistics and their expected values
is quantified by the average divergence:
\begin{equation}\label{eq:obj}
  \phi(\theta)
  =
  \sum_{q=1}^Q
  w_q
  \
  D\left( \langle \hat S_y \rangle_q,  \langle S_y \rangle_q \right) 
\end{equation}
where the positive weight $w_q$ is (as above) proportional to the size
of the $q$-th spectral domain and where $D(\cdot,\cdot)$ is a measure
of divergence between two $m\times m$ positive matrices defined as
\begin{equation}
  \label{eq:kullR}
  D(R_1,R_2) 
  =
  \trace \left( R_1R_2\inv \right) - \log\det (R_1R_2\inv) - m
\end{equation}
This is nothing but the Kullback divergence between two
$n$-dimensional zero-mean Gaussian distributions with positive
covariance matrices $R_1$ and $R_2$ respectively.

The reason for using the mismatch measure~(\ref{eq:obj}) is its
connection to maximum likelihood principle (see below).
Even though the divergence~(\ref{eq:obj}) may, at first sight, seem
more difficult to deal with than a simple quadratic distance, it is
actually a better choice in at least two respects: first, we expect it
to yield efficient parameters estimates because of the asymptotic
optimality of maximum likelihood estimation; second, thanks to its
connection to the log-likelihood, it lends itself to simple
optimization via the EM algorithm (see below).

A last note: since domain averaging does not change the algebraic
structure of the spectral covariance matrices (\emph{i.e.}
eq.~(\ref{eq:specmody}) becomes~(\ref{eq:modbydom})), it does not
introduce any bias in the estimation of $A$.

\section{MAXIMUM LIKELIHOOD AND EM}

\subsection{Whittle approximation}\label{sec:whittle}

The Whittle approximation is a spectral approximation to the
likelihood of stationary processes.
It has been introduced for the blind separation of noise-free mixtures
by Pham~\cite{pham:eusipco2000}.  Simplifying a little bit, this
approximation boils down to asserting that the coefficients $Y(\nu)$
of definition~(\ref{eq:dft}) taken at frequencies $\nu =p/T$ are
uncorrelated, have zero-mean and a covariance matrix equal to
$S_y(\nu)$.
Simple computations then show that ---up to a constant and a scalar
factor--- the (negative) log-likelihood of the data takes the
form~(\ref{eq:obj}) under the additional approximation that
$S_y(\nu )$ is constant over each spectral domain.

The Whittle approximation is good for Gaussian processes but certainly
does not capture all the probability structure, even for large $T$,
for non Gaussian processes.  However, it it still provides a
principled way of exploiting the spectral structure of the process,
leading to the selecting of~(\ref{eq:obj}) as an objective function.
In addition, it suggests to use the EM algorithm for
minimizing~(\ref{eq:obj}).

\subsection{An EM algorithm in the spectral domain}\label{sec:domain-em}

Using the EM technique for maximizing a likelihood function requires
defining latent (unobserved) data.  In the case of source separation,
there is an obvious choice: take the components as the latent data.
This approach was introduced in~\cite{EMMG:icassp97} for a noisy non
Gaussian i.i.d. model of source separation and later
in~\cite{snoussi:MaxEnt2001} for a noisy non Gaussian model in the
spectral domain.  Both these models lead to heavy computation.  In
contrast, by i) using only a Gaussian model (the Whittle
approximation) and ii) averaging over spectral domains, the EM
algorithm becomes very computationally attractive.

Room is lacking for a complete derivation of the EM algorithm in our
case but it is not difficult to adapt, for instance, the computations
of~\cite{EMMG:icassp97} to our case.  Let us only mention why the
resulting algorithm is much simpler.

First, when dealing with data structured as $y=As+n$, EM needs to
evaluate conditional expectations $E(s|y)$ and $E(ss\adj|y)$.  Thanks
to the Gaussian model, these are readily found to be \emph{linear}
functions of $y$ and $yy\adj$ respectively:
\begin{eqnarray}
  \label{eq:esycy} E(s|y) &=& Wy
  \\
  \label{eq:esscy} E(ss\adj|y) &=& Wyy\adj W\adj + C 
\end{eqnarray}
where matrices  $C$ and $W$ are defined as
\begin{eqnarray}
  \label{eq:defC} C &=& (A\adj R_n\inv A + R_s\inv )\inv \\
  \label{eq:defW} W &=& (A\adj R_n\inv A + R_s\inv )\inv  A\adj R_n\inv 
\end{eqnarray}
with covariance matrices $R_s =\mathrm{Cov}(s)$, $R_n
=\mathrm{Cov}(n)$.

Second, this linearity is preserved through domain averaging, meaning
that the EM algorithm only needs to operate on the sample covariance
matrices $\langle \hat S_y \rangle_q$.  This set of matrices is a
sufficient statistic set in our model; it is also all that is needed
to run the EM algorithm.

Thus, blind separation of noisy mixtures of stationary sources can be
achieved by computing the periodogram, averaging it into a set of
sample covariance matrices and maximizing the likelihood by then
running the EM algorithm.  The algorithm is summarized in pseudo-code
(see Alg.~\ref{tab:em}), but its derivation (which is purely
mechanical) is omitted.
\floatname{algorithm}{Alg.}
\begin{algorithm}
  \caption{Gaussian EM algorithm over spectral domains}  \label{tab:em}
  \begin{algorithmic}[1]
    \STATE \textbf{Start} with sample covariance matrices  $\langle \hat S_y \rangle_q$, 
    and initial guesses for $A$, $S_n$ and $\langle S_s\rangle_q$.
    \REPEAT
    \STATE \COMMENT {E-step \hrulefill Compute conditional statistics}
    \FOR {$q=1$ to $Q$} 
    \STATE $C_q       = ( A\adj S_n \inv A + \langle S_s \rangle_q \inv )\inv $
    \STATE $R_{yy}(q) = \langle \hat S_y \rangle_q  $ 
    \STATE $R_{ys}(q) = \langle \hat S_y \rangle_q   S_n\inv A\  C_q $
    \STATE $R_{ss}(q) =  C_q \ A \adj S_n \inv \langle \hat S_y \rangle_q  S_n \inv A\  C_q + C_q$
    \ENDFOR
    \STATE $R_{ss} = \frac 1 T \sum_{q=1}^Q\ w_q  R_{ss} (q)   $
    \STATE $R_{ys} = \frac 1 T \sum_{q=1}^Q\ w_q  R_{ys} (q)   $
    \STATE $R_{yy} = \frac 1 T \sum_{q=1}^Q\ w_q  R_{yy} (q)   $
    \STATE \COMMENT {M-step \hrulefill Update the parameters}
    \STATE $A    = R_{ys}R_{ss}\inv$ 
    \STATE $S_n  = {\diag} \left( R_{yy} - R_{ys}  R_{ss}\inv  R_{ys}\adj \right) $
    \STATE $\langle S_s\rangle_q = \diag \left( R_{ss}(q) \right) $ for $1\leq q \leq Q$.
    \STATE Renormalize $A$ and the $\langle S_s\rangle_q$
    \UNTIL {convergence}
  \end{algorithmic}
\end{algorithm}
In this pseudo-code, the $\mathrm{diag}(\cdot)$ operator sets to $0$
the off-diagonal elements of its argument.  We also include a
renormalization step which deals with the scale indetermination
inherent to source separation: each column of $A$ is normalized to
have unit norm and the corresponding scale factor is applied to the
average source spectra.

\section{APPLICATION AND COMMENTS}

\subsection{Separating astrophysical components}
Preliminary tests have been carried on simulated observations in six
channels at microwave frequencies 100, 143, 217, 353, 545 and 857 GHz,
over sky patches of size $12.5^\circ \times 12.5^\circ$ sampled over a
$300 \times 300$ pixel grid.  Mixtures include three astrophysical
components (CMB, Galactic dust, and emission from galaxy clusters) and
white noise.

Since isotropic observations are expected, we choose spectral domains
which are not only symmetric but also rotation invariant; in other
words: spectral rings.
The sample spectral covariance are computed over 15 such rings equally
spaced over the whole band.  Data reduction thus is by a factor of
$(6\times 300\times 300)/ (15\times 6 \times 6) = 1000$.  The EM
algorithm converges in a few tens of iterations amounting to a few
seconds on a 1 GHZ machine when coded in octave (a free clone of
Matlab: \texttt{http://www.octave.org}).

\begin{figure}[htbp]
  \label{fig:exp}
  \begin{center}
    \includegraphics[width=\columnwidth]{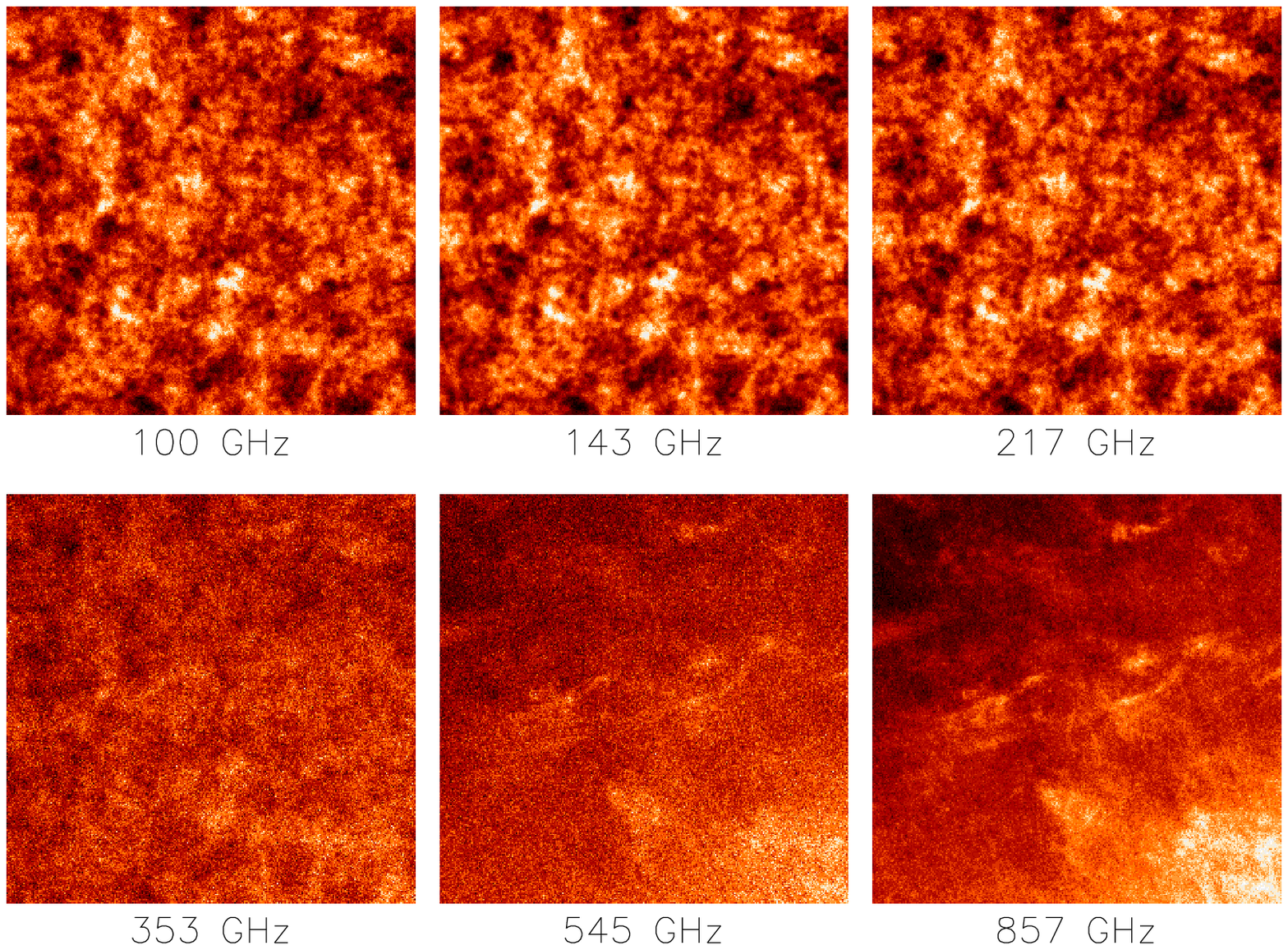}    
    \includegraphics[width=\columnwidth]{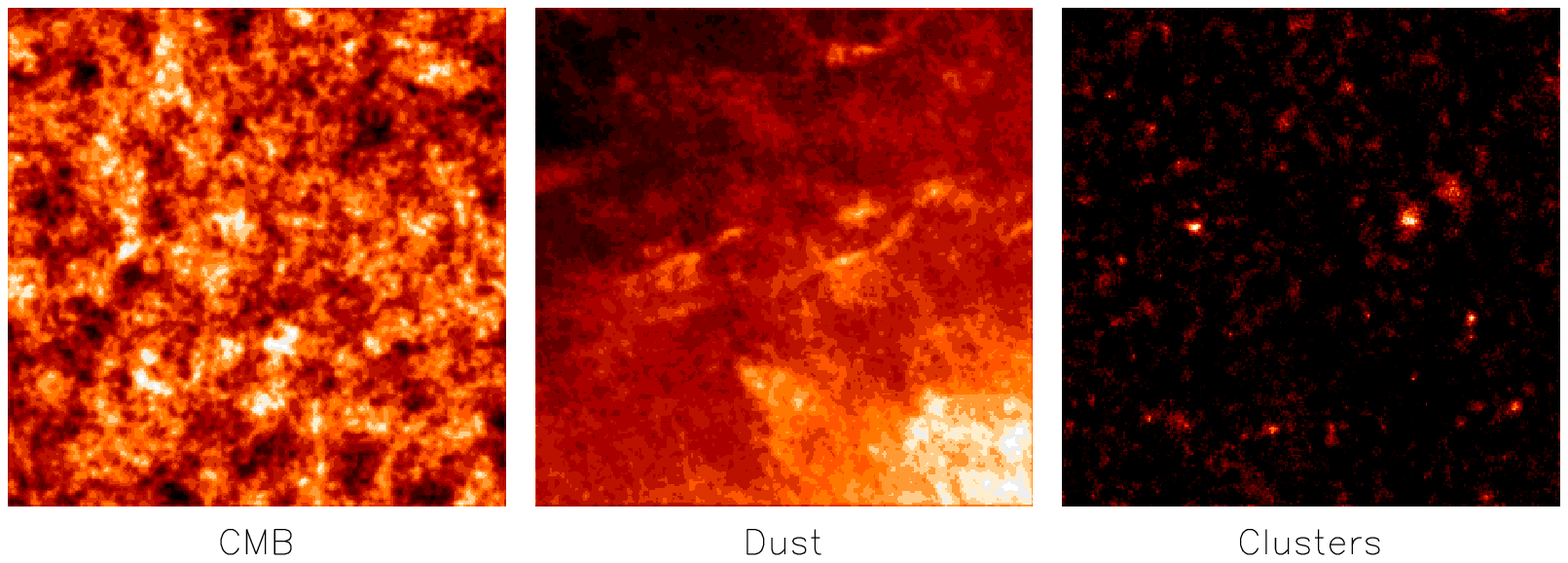}  
    \caption{Top two rows: maps at the detectors.  
      Bottom row: components extracted with the Wiener filter based on
      the estimated parameters.}
  \end{center}
\end{figure}

Room is lacking for a detailed description of our experiments which
will be reported elsewhere.  See figure~\ref{fig:exp} for an
illustration with a typical (and significant) level of noise.
A notable feature here is the separation of the galaxy clusters.  The
SNR on the CMB component is also much improved at high frequency even
though this cannot be assessed from the picture.  See the companion
paper in these proceedings for more details.

\subsection{Conclusion}\label{sec:conclusion}

We have proposed an efficient method to maximize the likelihood of a
model of noisy mixtures of stationary sources by implementing the EM
algorithm on spectral domains.  The procedure jointly estimates all
the parameters: mixing matrix, average source spectra, noise level in
each sensor.  Spectral averaging offers large computational savings,
especially when dealing with images.

Since the inference principle is maximum likelihood for a `smooth'
Gaussian stationary model, we expect a good statistical efficiency
when the source spectra are reasonably smooth (even though we saw
little performance degradation in our experiments when using a very
coarse $Q=2$ spectral partition) and when the sources actually are
Gaussian.  In the CMB application, some components are very close to
Gaussian (the CMB itself) while others are strongly non Gaussian; it
is not clear yet how to best combine non Gaussian information with
spectral diversity.

As final note, we recall that, in the noise-free case, the ability to
blindly separate Gaussian stationary components rests on
\emph{spectral diversity}: the spectra of any two sources should not
be proportional.  The noisy case is complicated by the fact that the
noise parameters also have to be estimated.

Future research should cover many issues: blind identifiability in
unknown noise, choice of the spectral domains, integration of non
Gaussian information, integration of prior information,\ldots

\small{

}

\end{document}